\documentclass[nofootinbib,pra,twocolumn,a4paper,groupedaddress,showpacs,floatfix,aps,10pt,reprint]{revtex4-1}

\usepackage{graphicx,amsmath,amssymb,amsfonts,dsfont,subfigure,color}

\newcommand{\ket}[1]{\ensuremath{|#1\rangle}}
\newcommand{\mc}[1]{\ensuremath{\mathcal{#1}}}
\newcommand{\bra}[1]{\ensuremath{\langle #1 |}}

\newcommand{\vroW}{\varrho}

\begin{document}

\title{Spectral shift and dephasing of electromagnetically induced transparency in an interacting Rydberg gas}

\author{Jingshan Han,${}^1$ Thibault Vogt,${}^{1,2}$ and Wenhui Li${}^{1,3}$}

\affiliation{${}^1$Centre for Quantum Technologies, National University of Singapore, 3 Science Drive 2, Singapore 117543}
\affiliation{${}^2$MajuLab, CNRS-UNS-NUS-NTU International Joint Research Unit UMI 3654, Singapore 117543}
\affiliation{${}^3$Department of Physics, National University of Singapore, Singapore 117542}


% 42.50.Gy Effects of atomic coherence on propagation, absorption, and amplification of light; electromagnetically induced transparency and absorption
% 32.80.Ee Rydberg states

\begin{abstract}
We perform spectroscopic measurements of electromagnetically induced transparency (EIT) in a strongly interacting Rydberg gas. We observe a significant spectral shift and attenuation of the transparency resonance due to the presence of interactions between Rydberg atoms. We characterize the attenuation as the result of an effective dephasing, and show that the shift and the dephasing rate increase versus atomic density, probe Rabi frequency, and principal quantum number of Rydberg states. Moreover, we find that the spectral shift is reduced if the size of a Gaussian atomic cloud is increased, and that the dephasing rate increases with the EIT pulse duration at large parameter regime. We simulate our experiment with a semi-analytical model, which yields results in good agreement with our experimental data.
\end{abstract}

\maketitle

%
%%%%%%%%%%%%%%%%%%%%%%%%%%%%%%
% Introduction
%%%%%%%%%%%%%%%%%%%%%%%%%%%%%%
%
\section{Introduction \label{introduction}}
Atomic systems driven in the configuration of electromagnetically induced transparency involving a Rydberg state (Rydberg EIT) \cite{fleischhauer:05} are promising systems for the investigation of quantum nonlinear optics \cite{mohapatra:08,pritchard2010cooperative} and quantum many-body physics \citep{low2012experimental,weimer2010rydberg,schauss2012observation,labuhn2016tunable} with strong interactions and correlations. With Rydberg EIT, non-linearity at the single photon level has been demonstrated with the remarkable realizations of photon filters, deterministic single photon sources, and interaction between pairs of photons \citep{dudin2012strongly,peyronel2012quantum,firstenberg2013attractive}, and also been utilized to implement single-photon transistors \citep{gorniaczyk2014sptransistor,tiarks2014sptransistor}. Moreover, the recent demonstration of interaction enhanced absorption imaging (IEAI) \citep{gunter2012interaction,gunter2013observing} confirms the great potential of Rydberg EIT for the study of many-body physics with Rydberg atoms.

In Rydberg EIT, highly correlated many-body atomic states arising from the interaction-induced blockade \cite{jaksch:00,tong2004local,vogt2006dipole,heidemann2007strongblockade} result in strong nonlinear optical response, and can be mapped onto the probe field propagating through the medium. Thus detailed studies on the spectral, temporal, and spatial properties of the transmitted probe light are crucial for proper understanding of such systems and their further applications in quantum optics, quantum information, and quantum many-body physics. Because of the large parameter space and complex dynamics of Rydberg EIT ensembles in the blockade regime, many questions remain open in spite of significant investigation efforts, both experimental \cite{weatherill:08,pritchard2010cooperative,schempp2010cpt,sevinccli2011quantum,raitzsch2009dephasing,baluktsian2013evidence,zhang2014autler,desalvo2016rydberg} and theoretical \cite{weimer2008quantum,petrosyan2011electromagnetically,ates2011electromagnetically,sevinccli2011nonlocal,sevinccli2011quantum,heeg2012hybrid,garttner2012finite,PetrosyanCorrelations2013,GarrtnerSemianalytical2014}.
All of Rydberg EIT experiments in the blockade regime have observed the reduction of the EIT transparency at the probe resonance due to Rydberg blockade induced dissipation, also called photon blockade \cite{gorshkov:11}. However, the spectral shift of the transparency resonance, as predicted by several theoretical models, has only been observed in two experiments \cite{carr2013nonequilibrium,desalvo2016rydberg}, but remains elusive in others. Theoretical calculations, while succeeding to describe certain experimental observations, have yet converged to a comprehensive physical picture due to the challenges posted by strong correlation and dissipation in such systems.

In this article, we present a study of the Rydberg EIT spectra in presence of interaction and their dependence on atomic density, input probe light intensity, and principal quantum number of Rydberg states. We show that all the features of the spectra can be quantified phenomenologically via an effective shift and an effective dephasing rate in the optical Bloch equations describing the evolution of non interacting atoms driven under conditions of Rydberg EIT, as in reference \cite{raitzsch2009dephasing}.
Our experimental data are in good agreement with a semi-analytical model based on previous results obtained from the Monte Carlo rate equations approach \cite{ates2011electromagnetically} and the superatom model \cite{petrosyan2011electromagnetically}. We find that the spectral shift is accurately accounted for by an average energy level shift due to the interaction between Rydberg atoms, while the effective dephasing rate mainly comes from the photon blockade effect and dephasing inherent to the dispersion of Rydberg energy level shifts. The dephasing rate increases with the EIT pulse duration at large parameter regime and deviates from the model because of the presence of additional dephasing mechanisms. Finally, we demonstrate that as the size of a Gaussian atomic cloud is increased, the measured spectral shift is reduced. This study provides important experimental evidences to advance the understanding on the spectral properties of Rydberg EIT.

%
%%%%%%%%%%%%%%%%%%%%%%%%%%%%%
% main body with results
%%%%%%%%%%%%%%%%%%%%%%%%%%%%%
\section{Experiment \label{setup}}
\begin{figure}[hptb]
\includegraphics[width=8cm]{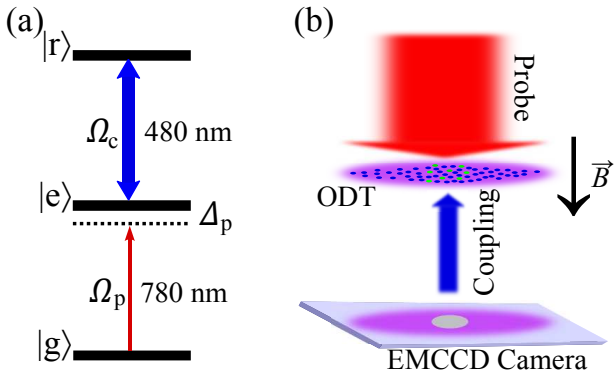}%%
\caption { (color online) (a) Three-level Rydberg EIT scheme for spectroscopic measurements. A probe beam of Rabi frequency $\Omega_p$ is detuned $\Delta_p$ from the $|5s_{1/2}, F=2, m_F=2\rangle$ ($|g\rangle$) $\rightarrow$ $|5p_{3/2}, F=3, m_F=3\rangle$ ($|e\rangle$) transition and a coupling beam of Rabi frequency $\Omega_c$ is on resonance with the $|5p_{3/2}, F=3, m_F=3\rangle$ $\rightarrow$ $|n s_{1/2}, J=1/2, m_J=1/2\rangle$ ($|r\rangle$) transition, where $n$ is the principal quantum number of a Rydberg state. (b) Schematics of the experimental configuration. The probe and coupling beams counter-propagate coaxially along the $z$ direction in the $\sigma^+$-$\sigma^-$ polarization configuration. While the probe beam has a collimated $1/e^2$ Gaussian radius of 3.45 mm, the coupling beam is focused at the center of the atomic cloud with a $1/e^2$ Gaussian radius of 50 $\mu$m and a peak Rabi frequency of $\Omega_{c0}$. The transparent spot in the shadow of the atomic cloud on the EMCCD (electron-multiplying charge-coupled device) camera screen illustrates the transparency window opened by the coupling beam for the probe beam to pass through. \label{ExperimentSetup}}
\end{figure}

The experiment was performed with a $^{87}$Rb atomic cloud released from a horizontally positioned single-beam optical dipole trap (ODT), which was formed by a 1064 nm laser beam with a power of 3.1 W and a $1/e^2$ Gaussian radius of 42.6 $\mathrm{\mu m}$. Atoms were loaded into the ODT from a molasses cooled atomic ensemble, the preparation of which was detailed in Ref.\cite{han2015lensing}. Subsequently, a guiding field of approximately 3.5 Gauss along the vertical direction pointing downwards was switched on to define the quantization axis $z$, and the atoms in the ODT were optically pumped into the $|5s_{1/2}, F=2, m_F=2\rangle$ ($|g\rangle$) state. At this stage, the atomic cloud had a temperature in the range 20 to $40\ \mu$K. The atomic cloud was then released for a time of flight (TOF) before the spectroscopic measurement. By changing the ODT loading efficiency and/or the TOF duration, the peak atomic density of the ground state $|g\rangle$, $n_0$, and the $1/e^2$ radius of the cloud along the vertical direction, $w_z$, could be varied independently in the ranges of $0.1 - 3 \times10^{11}\  \mathrm{cm^{-3}}$ and  $15 - 80\ \mathrm{\mu m}$, respectively.

Shown in Fig.~\ref{ExperimentSetup} are the schematics of the energy levels and the optical setup for Rydberg EIT, the details of which are similar to that in Ref.\cite{han2015lensing}. At each experimental cycle after TOF of the atomic cloud, the probe and coupling beams were turned on simultaneously for 15 $\mu$s during which an EMCCD camera was exposed to acquire the image of the probe beam transmitted through the atomic cloud via a diffraction limited optical system. To obtain an EIT transmission spectrum, a set of images of the transmitted probe light were taken while varying the probe beam detuning $\Delta_p$ from shot to shot to scan through the resonance of the $\ket{g}\leftrightarrow\ket{e}$ transition but fixing the coupling beam frequency to be on resonance with the $\ket{e}\leftrightarrow\ket{r}$ transition. The probe transmission ($I/I_0$) was extracted by taking the ratio between the probe intensity $I$ passing through the center of the coupling beam (with Rabi frequency $\Omega_{c0}$) and that of the incoming probe beam without atomic cloud ($I_0$), and plotted vs. $\Delta_p$ to get the transmission spectra in Fig.~\ref{ExperimentPlot}.

%%%%%%%%%%%%%%%%%%%%%%%%%%%%%%
% Results and Discussion
%%%%%%%%%%%%%%%%%%%%%%%%%%%%%
\section{Results and discussion \label{results}}

In EIT of a non-interacting gas, the linear susceptibility for the probe light at the first order is given by \cite{fleischhauer:05}
\begin{equation}	
\chi^{(1)} \left ( \vec r \right) =i \frac{n_{at}\left ( \vec r \right) \Gamma_{e} \sigma_0 \lambda}{ 4 \pi \left(\gamma _{ge}-i \Delta _p+\frac{\Omega _c \left ( \vec r \right) ^2}{4 (\gamma _{gr}-i (\Delta _c+\Delta _p))}\right)} \label{susceptibility},
\end{equation}
where $\lambda$ is the wavelength of the probe transition, $\sigma_0=3 \lambda^2 /2 \pi$ is the resonant cross-section of the probe transition,  $\Gamma_{e}=2\pi\times6.067$ MHz is the decay rate of intermediate state $|e\rangle$, $\Delta _p$ and $\Delta _c$ are the detunings of the probe and coupling lights, and finally $\gamma _{ge}\approx\Gamma_e/2$ and $\gamma _{gr}$ are the decay rates of atomic coherences. As the atomic sample in this experiment is rather thin along the beam propagation direction $z$, lensing can be neglected \cite{han2015lensing}, and the spatial dependent terms in Eq.~(\ref{susceptibility}) are reduced to a constant $\Omega _{c0}$ and one-dimensional functions $\chi^{(1)} \left ( z \right)$ and $n_{at}\left ( z \right)$. As the incoming probe light of Rabi frequency $\Omega_{p0}$ propagates through the atomic sample, the solution of the one-dimensional Maxwell's equation $\partial_{z} \Omega_{p} =   \frac {i k} {2} \chi^{(1)} \Omega_{p}$ gives the transmission
\begin{equation}
T(\Delta_p, \gamma_{gr},\Delta_c,OD)=\exp \left( -OD \times \mathrm{Im} \left[ \frac{k \chi^{(1)}}{n_{at} \sigma_0} \right] \right),
\label{fitformula}
\end{equation}
where $k$ is the wavenumber of the probe light $k = 2\pi / \lambda$ and $OD$ corresponds to the optical density of the atomic cloud at the first order.

In an interacting Rydberg gas, the single-atom Rydberg energy level is shifted by the strong interaction, which for $ns$ state in our experiment is the repulsive van der Waals interaction $V(r) = - C_6/r^6$, where the strength coefficient $C_6$ scales as $n^{11}$ \cite{singer2005long}. One of the most important consequences of this energy level shift is the blockade effect that allows only one Rydberg excitation among many atoms within a sphere of radius $R_B$. The interaction induced level shift and excitation blockade greatly modify the susceptibility of the EIT ensemble. The linear susceptibility in Eq.~(\ref{susceptibility}) is now changed to a nonlocal nonlinear susceptibility, which in principle is given by the solution of a full master equation of the interacting many-body system, and this change is reflected in the EIT transmission spectra. In a simple and intuitive physical picture, the shifted Rydberg energy level detunes the transparency away from that of the non-interacting EIT resonance, and the blockade effect results in scattering around each blockaded sphere, which gives rise to additional decoherence compared to that of the single-atom EIT. The resulting spectral shift and attenuation of the transparency peak depend on the interaction strength $C_6$, the density of blockaded spheres (Rydberg excitations) $n_{Ryd}$, and the number of atoms inside each blockaded sphere $N_B$.

\begin{figure}[hptb]
\includegraphics[width=8.3cm]{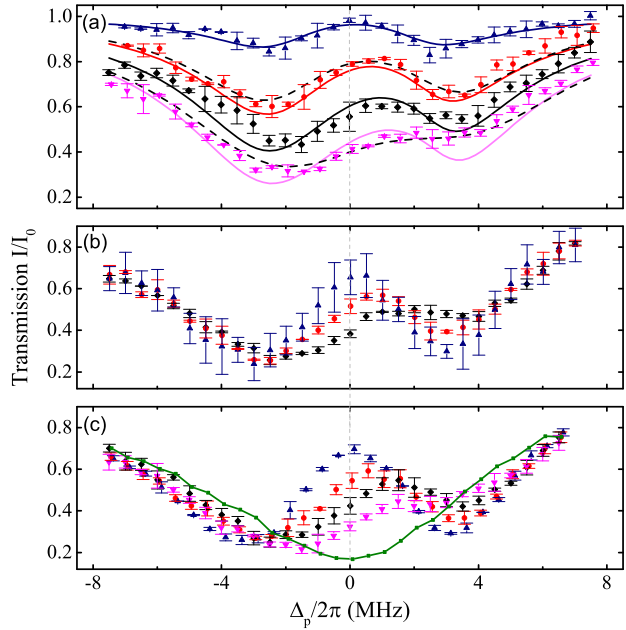}%
\caption {(color online) Transmission spectra of the probe light. The spectra are taken at (a) $\Omega_{c0}/2\pi = 5.2\ \mathrm{MHz}$, $\Omega_{p0}/2\pi = 1.45\ \mathrm{MHz}$, $w_z = 18\ \mu m$, and Rydberg state $n s = 38 s$ with different atomic densities, $n_0 = 0.30\times10^{11}\ \mathrm{cm}^{-3}$ ($\blacktriangle$), $1.17\times10^{11}\ \mathrm{cm}^{-3}$ ($\bullet$), $1.82\times10^{11}\ \mathrm{cm}^{-3}$ ($\blacklozenge$), and $2.64\times10^{11}\ \mathrm{cm}^{-3}$ ($\blacktriangledown$), respectively; (b) $\Omega_{c0}/2\pi = 5.6\ \mathrm{MHz}$, $n_0 = 2.14 \times10^{11}\ \mathrm{cm}^{-3}$, $w_z = 21\ \mu m$, and $n s = 38 s$ with different incoming probe Rabi frequencies $\Omega_{p0}/2\pi = 0.68\ \mathrm{MHz}$ ($\blacktriangle$), $1.25\ \mathrm{MHz}$ ($\bullet$), and $1.8\ \mathrm{MHz}$ ($\blacklozenge$), respectively;  (c) $\Omega_{c0}/2\pi = 5.6\ \mathrm{MHz}$, $n_0 = 2.14\ \times10^{11}\ \mathrm{cm}^{-3}$, $w_z = 21\ \mu m$, and $\Omega_{p0}/2\pi = 1.45\ \mathrm{MHz}$ with different Rydberg states, $27 s$ ($\blacktriangle$), $33 s$ ($\bullet$), $38 s$ ($\blacklozenge$) and $43 s$ ($\blacktriangledown$), respectively. The uncertainties for the experimental parameters are less than $10\%$ for $n_0$ and $w_z$, and less than $3\% $ for $\Omega_{c0}$ and $\Omega_{p0}$. Each spectrum is an average of 3 or more scans. In (a), the dashed lines correspond to fittings of experimental data to Eq.~(\ref{fitformula}), whereas the solid lines are generated by the theoretical model with the corresponding experimental parameters as inputs (see text). In (c) the solid line is the two-level absorption curve of the probe light in the absence of the coupling light. \label{ExperimentPlot}}
\end{figure}

A set of values around the middle of our experimental parameter range, $ns = 38 s$, $n_0 = 2.1\times10^{11}\ \mathrm{cm}^{-3}$, $\Omega_{c0}/2\pi = 5.6\ \mathrm{MHz}$, $\Omega_{p0}/2\pi = 1.45\ \mathrm{MHz}$, $\Delta_p = 0$, gives the blockade radius $R_B = \left(-2 C_6/\gamma_{\mathrm{EIT}}\right)^{1/6} $ = 2.4 $\mu$m, the atom number per unit blockaded sphere $N_B = n_0 \times 4 \pi R_{B}^3/3 = 5$, and the Rydberg atom density $n_{Ryd} = n_0 \times f_R = 1.05\times10^{10}\ \mathrm{cm}^{-3}$, where $\gamma_{\mathrm{EIT}} = \Omega_{c0}^2/ \Gamma_e$ is the on resonance single atom EIT linewidth \cite{peyronel2012quantum} and $f_R$ is the Rydberg excitation fraction. This indicates that the Rydberg EIT system in our experiment goes into the blockade regime of this interacting many-body system. In the three sets of transmission spectra in Fig.~\ref{ExperimentPlot}, there is an increasing blue shift of transparency away from the single atom EIT resonance ($\Delta_p = 0$) as well as an increasing width and reduced height of the transparency, when the ground state atomic density $n_0$, the incoming probe Rabi frequency $\Omega_{p0}$, or the principal quantum number $n$ are increased.

\begin{figure}[hptb]
\includegraphics[width=8.7cm]{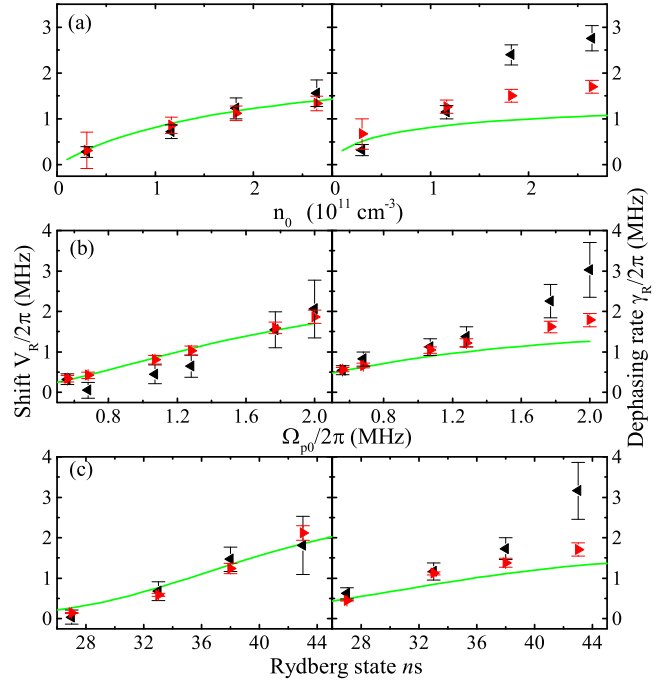}%
\caption {(Color online) Spectral shift and dephasing rate of EIT transmission spectra. The spectral shift $V_R=-\Delta_c$ and dephasing rate $\gamma_R=\gamma_{gr}$ are extracted by fitting the experimental($\blacktriangleleft$) and the simulated ($\blacktriangleright$) transmission spectra to Eq.~(\ref{fitformula}), and plotted against (a) $n_0$, (b) $\Omega_{p0}$, and (c) $n$. The solid lines are plots of $\Delta_R$ and $\sqrt{\vartheta_R}$ as expressed in Eqs.~(\ref{meanfield}) and ~(\ref{dispersion}) with $n_{at} = n_0$ and $\Omega_{p} = \Omega_{p0}$. For these measurements done in the thin atomic samples, this simple expression of $\Delta_R$ seems to agree well with the observed shift, while $\sqrt{\vartheta_R}$ only gives a good account of the experimental data at very small dephasing rates.
\label{ExtractedPlotEMF}}
\end{figure}

In order to quantify the spectral shift $V_R$ and the attenuation of the transparency peak, we fit the experimental EIT transmission spectra to Eq.~(\ref{fitformula}) with $\Delta_c$, $\gamma_{gr}$ and $OD$ as the fitting parameters. While experimentally the coupling light frequency is always on the resonance of the single-atom transition $\ket{e}\leftrightarrow\ket{r}$, $\Delta_c$ is set to be a fitting parameter to account for the shift $V_R=-\Delta_c$ due to Rydberg interactions. The effective dephasing rate $\gamma_{R} = \gamma_{gr}$ is also set to be a fitting parameter, and it accounts mostly for a reduction of the transparency peak with exponential dependence in conditions of relatively large $\Omega_c$ \cite{fleischhauer:05}. The effective dephasing with rate $\gamma_{R}$ includes the decay of atomic coherence for the single-atom Rydberg EIT, which has a rate of $\gamma_0 \approx 2 \pi \times 100$ kHz in our experiment \cite{han2015lensing}, the effect of the photon blockade, and other dephasing processes, such as the dephasing inherent to the dispersion of the shifted Rydberg energy levels due to interaction. The result of the fits captures essentially all the features of the spectra, as illustrated in Fig.~\ref{ExperimentPlot} (a) with a couple sample fittings. As shown in Fig.~\ref{ExtractedPlotEMF}, the extracted shift $V_R=-\Delta_c$ and dephasing rate $\gamma_{R} = \gamma_{gr}$ are increasing with $n_0$, $\Omega_{p0}$, and $n$, and their amplitudes are of the same order.

In order to get a deeper insight on the different contributions to the effective dephasing rate, we simulate our experimental transmission spectra with an approximate  nonlinear susceptibility $\overline{\chi}$ \cite{petrosyan2011electromagnetically,GarrtnerSemianalytical2014},
\begin{equation}
\overline{\chi} = \chi_{B} f_R \left(N_B-1\right)+\chi_E \left( 1-f_R (N_B -1)\right),
\label{approx_susceptibility}
\end{equation}
where $f_R (N_B-1)$ is the fraction of atoms inside the blockaded spheres excluding Rydberg atoms and every physical quantity in the equation is a function of $z$. The first term is proportional to the average three-level susceptibility of atoms inside the blockaded spheres $\chi_{B}=\frac{1}{V_B}\int_0^{R_B} 4 \pi r^2 \chi\left(\Delta_c = C_6/r^6, \gamma_{gr}=\gamma_0\right) dr$. In this integral and as detailed in Appendix \ref{ThreeLevel}, $\chi\left(\Delta_c = C_6/r^6, \gamma_{gr}=\gamma_0 \right)$ is the three-level atom susceptibility calculated from the steady state solution $\vroW$ of the single atom master equation, computed at fourth order in $\Omega_p$ for $\Delta_c = C_6/r^6$ and $\gamma_{gr}=\gamma_0$. The blockade radius $R_B$ is defined as the distance at which $\vroW_{rr}(\Delta_c=C_6/R_B^6)=\vroW_{rr}(\Delta_c=0)/2$, where $\vroW_{rr}(\Delta_c)$ is the Rydberg population obtained from the single atom master equation at given detuning $\Delta_c$, and finally $V_B=\frac{4}{3} \pi R_B^3$ is the volume of the blockaded sphere. Because of the very large interaction with the Rydberg atom at the center of the blockade sphere, $\chi_B$ is nearly equal to the two-level atom susceptibility $\chi(\Delta_c= -\infty, \gamma_{gr}=\gamma_0)$, and therefore attributes mostly to photon blockade. $\chi_E$ in the second term of Eq. \eqref{approx_susceptibility} is the susceptibility of unblockaded atoms (including Rydberg excitations themselves) more than $R_B$ distance from any (other) Rydberg excitations. Since the unblockaded atoms could be excited to interact with other Rydberg excitations, $\chi_E = \chi\left(\Delta_c = -\Delta_R, \gamma_{gr} = \sqrt{\vartheta_R} + \gamma_0\right)$ is the susceptibility of three-level Rydberg EIT with the shifted energy level, where $\Delta_R$ accounts for the average energy level shift due to interactions with surrounding Rydberg atoms, while $\sqrt{\vartheta_R}$ accounts for the standard deviation of the level shift from the mean value $\Delta_R$.\footnote{We assume that the bandwidth of energy level shifts is of the order of $2 \sqrt{\vartheta_R}$. The effect of this bandwidth is simulated with a dephasing term $\sim \sqrt{\vartheta_R}$ in $\chi_E$, following an approach very similar to that used in optical Bloch equations for the treatment of non-zero laser linewidths \cite{gea1995electromagnetically}, although here the bandwidth of the distribution of energy level shifts is not strictly speaking of Lorentzian type.} $\Delta_R$ and $\sqrt{\vartheta_R}$ are evaluated with a simple mean-field approach \cite{weimer2008quantum,desalvo2016rydberg}, where the average shift felt by an atom $i$ is the sum over the interactions with all Rydberg excitations $j$ outside a sphere of radius $R_B$, $\Delta_R\approx \overline{\sum_j \frac{-C_6}{R_{ij}^6}}$, and the variance of this average shift is given by $\vartheta_R=\overline{\left(\sum \frac{-C_6}{R_{ij}^6}\right)^2}-\Delta_R^2$. At each spatial position $z$, $\Delta_R$ and $\vartheta_R$ are calculated\footnote{Correlations in position between Rydberg excitations are neglected in this evaluation of the variance of energy level shifts.} by replacing the sum with an integral and using the local density approximation (LDA) for $n_{at}(z)$ and $\Omega_p (z)$, which yields:
\begin{equation}
\Delta_R \approx \int_{R_B}^{\infty} f_R  n_{at}   \frac{-C_6}{r^6} 4 \pi r^2 dr= - \frac{4  \pi C_6 f_R n_{at} }{3 R_B^3},
\label{meanfield}
\end{equation}
and
\begin{equation}
\vartheta_R \approx \int_{R_B}^{\infty} f_R n_{at}   \frac{C_6^2}{r^{12}} 4 \pi r^2 dr=\frac{4  \pi C_6^2 f_R n_{at}}{9 R_B^9}.
\label{dispersion}
\end{equation}
Here the excitation fraction $f_R$ in the blockade regime is given in good approximation by
$f_R=\frac{f_0}{1-f_0+f_0 n_{at} V_B}$ \cite{GarrtnerSemianalytical2014,PetrosyanCorrelations2013}, where $f_0 = \vroW_{rr}(\Delta_c=0)$ is the Rydberg population fraction of the non-interacting Rydberg EIT.

The simulated EIT spectra are produced by numerically solving the Maxwell's equation with the approximate susceptibility in Eq.~(\ref{approx_susceptibility}) and the experimental parameters as inputs, without any fitting parameter. They agree reasonably well with the experimental spectra, as shown by the samples in Fig. \ref{ExperimentPlot} (a). The shift $V_R$ and dephasing  rate $\gamma_R$ of these simulated spectra are extracted by fitting to Eq.~(\ref{fitformula}) as in the case of experimental spectra, and plotted along with the experimental results in Fig.~\ref{ExtractedPlotEMF}. The spectral shifts are definitely well captured by our model, and the dephasing rates deviate only at large $n_0$, $\Omega_{p0}$, or $n$.

\begin{figure}[hptb]
\includegraphics[width=8.4cm]{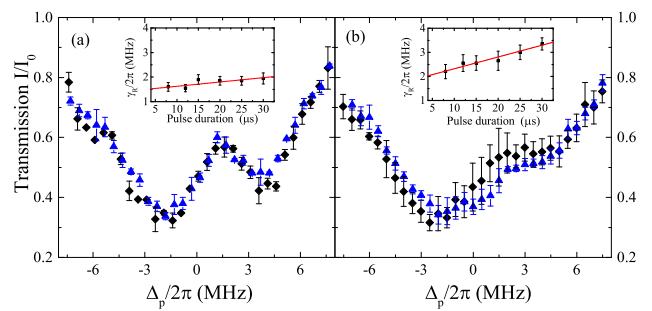}%
\caption {(color online) Time dependence of the dephasing rate. Transmission spectra are recorded with two different probe pulse durations, 8~$\mu$s ($\blacklozenge$) and 30~$\mu$s ($\blacktriangle$), for two different states, $ns=38s$ (a) and $ns=43s$ (b), respectively. The experimental parameters are $\Omega_{p0}/ 2 \pi=1.45$~MHz, $\Omega_{c0}/2 \pi=5.4$~MHz and $n_0=2.10\times10^{11}\ \mathrm{cm}^{-3}$ for $ns=38s$ and $n_0=2.60\times10^{11}\ \mathrm{cm}^{-3}$ for $ns=43s$, the uncertainties of which are the same as that in the caption of Fig.~\ref{ExperimentPlot}.  The dephasing rates are extracted  from  fittings to Eq.~(\ref{fitformula}) and  plotted versus pulse duration for each state, as shown in the insets. Solid lines are the results of linear regressions with slope coefficient 0.018(6)~MHz/$\mu$s for $ns=38s$ and 0.049(5)~MHz/$\mu$s for $ns=43s$, respectively.
\label{FigDephasing}}
\end{figure}

To further investigate the additional dephasing rates that are not included in our model, we perform spectroscopic measurements with different EIT pulse durations for two principal quantum numbers, as shown in Fig.~\ref{FigDephasing}. The dephasing rate becomes larger with increasing pulse duration, while the spectral shift remains the same. More measurements show that the time-dependent dephasing effect only becomes important at large parameter regimes, where our model deviates from the experimental data. One potential explanation is the presence of motional dephasing in our system, which could come from thermal motion of ground state atoms, photon recoil motion, and most likely dipole force induced motion \cite{baur2014single}.

\begin{figure}[hptb]
\includegraphics[width=8.5cm]{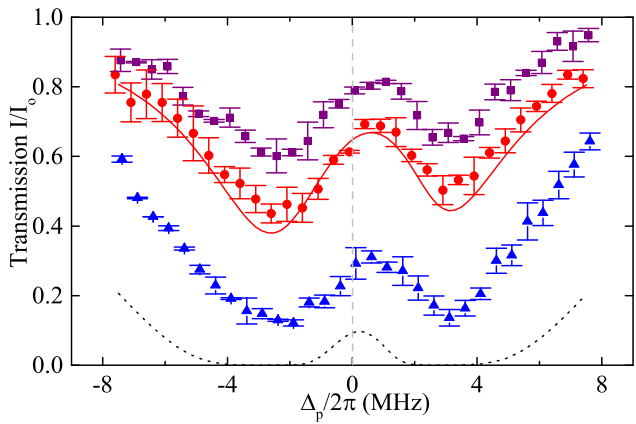}%
\caption { (color online) EIT transmission spectra obtained for different atomic cloud sizes, $w_z=17\ \mu$m ($\blacksquare$), $w_z=30\ \mu$m ($\bullet$) and $w_z=66\ \mu$m ($\blacktriangle$), respectively, with experimental conditions $\Omega_{c0}/2\pi = 5.2\ \mathrm{MHz}$, $\Omega_{p0}/2\pi = 1.5\ \mathrm{MHz}$, $n_0 = 1.2 \times10^{11}\ \mathrm{cm}^{-3}$, and  $ns = 38 s$. The uncertainties for the experimental parameters are the same as that in the caption of Fig.~\ref{ExperimentPlot}. Shown as well are the simulated spectral curves with the same experimental conditions and with the atomic cloud sizes of $w_z=30\ \mu$m  (solid line) and $w_z=198\ \mu$m (dotted line), the latter of which can not be realized in our current experimental configuration.
\label{SizeDependent}}
\end{figure}

An important experimental observation is that the spectral shift is smaller for a Gaussian atomic sample with larger size than for a smaller size one, while both have the same  peak atomic density, as shown in Fig.~\ref{SizeDependent}. This reduction in the observed shift can be explained by the inhomogeneity of the EIT ensemble together with the dependence shown in Fig.~\ref{ExtractedPlotEMF}. As the probe light enters the atomic sample, it first sees the lower atomic density at the wing part of the Gaussian density profile; and when the light reaches the center of the sample with the high atomic density, its intensity has already been attenuated due to scattering along its way. Given the similar peak atomic density, this effect of inhomogeneity in a thicker atomic sample obviously is more pronounced than that in a thinner one, and correspondingly results in smaller spectral shift.

%%%%%%%%%%%%%%%%%%%%%%%%%%%%%%
% conclusion
%%%%%%%%%%%%%%%%%%%%%%%%%%%%%
\section{Conclusion \label{conclusion}}
In conclusion, we have measured the spectral shift and dephasing rate of Rydberg EIT and clearly mapped out their dependence on density, probe Rabi frequency, and Rydberg principal quantum number. While the spectral shifts agree very well with theoretical predictions obtained in the frozen gas approximation over the whole parameter range, the dephasing rates are consistently larger than predicted at high Rydberg excitation or large blockade. This discrepancy is likely due to motional dephasing and invites further investigations. Furthermore, we demonstrate that the observed spectral shift also depends on the size of the Gaussian atomic sample. This clear observation of spectral shift raises the prospect to realize interaction induced optical switches, and these spectroscopic measurements are highly relevant for imaging Rydberg excitations via IEAI based on Rydberg EIT. Moreover, it would be interesting to extend our experiment into the parameter regime, where collective excitations could give a Rydberg excitation fraction higher than that of a noninteracting Rydberg gas \cite{garttner2014collective}.

\begin{acknowledgments}
The authors thank Beno\^{\i}t Gr\'{e}maud for helpful discussions. The authors acknowledge the support by the National Research Foundation, Prime Ministers Office, Singapore and the Ministry of Education, Singapore under the Research Centres of Excellence programme. This work is partly supported by Singapore Ministry of Education Academic Research Fund Tier 2 (Grant No. MOE2015-T2-1-085).
\end{acknowledgments}

\appendix
\section{}
\label{ThreeLevel}

The susceptibility $\chi$ for the probe light, obtained from the optical Bloch equations for three-level non-interacting atoms driven in configuration of EIT, is given by \cite{steck2006quantum}

\begin{align}
\chi = -\frac{2 n_{at} |d_{eg}|^2}{\varepsilon_0 \hbar \Omega_p} \vroW_{eg},
\label{chi}
\end{align}
where $d_{eg}$ is the dipole matrix element for the $|g\rangle\rightarrow|e\rangle$ transition, $n_{at}$ is the atomic density, $\Omega_p=\frac{-d_{eg} E_p}{\hbar}$ is defined as the Rabi frequency of the circular probe field $E_p/\sqrt{2} \left[ \cos \left( \omega_p t \right) \, \hat{e}_x+ \sin \left( \omega_p t\right) \hat{e}_y \, \right]$, and $\vroW_{eg}$ is the atomic coherence of the $|e\rangle\rightarrow|g\rangle$ transition. $\vroW_{eg}$ is given by the solution $\vroW$ of the Markovian master equation
$\partial_t \vroW = - \frac{i}{\hbar} [ H , \vroW ] +\mc{L} \vroW$, where

\begin{align}
H  = & - \hbar \left(\Delta_p \ket{e}\bra{e}+ \Delta_r\ket{r}\bra{r}\right) \notag \\
 & + \frac{\hbar}{2}\left( \Omega_p \ket{e}\bra{g} + \Omega_c \ket{r} \bra{e} \,+\,\text{h.c.}\right), \notag
\end{align}

and

\begin{align}
\mc{L}\vroW \approx &          -  \frac{\Gamma_e}{2}
\left( \ket{e}\bra{e}  \vroW  + \vroW \ket{e}\bra{e}
- 2 \ket{g}\bra{e} \vroW \ket{e}\bra{g} \right) \notag  \\
& -  \gamma_{gr}
\left( \ket{g}\bra{g}  \vroW \ket{r}\bra{r}  + \ket{r}\bra{r} \vroW \ket{g}\bra{g}  \right).   \notag
\end{align}
\normalfont
Here, $\Delta_{p}$ and $\Delta_r=\Delta_p+\Delta_c$ are the probe and the two-photon detunings, respectively. The decay rate of state $\ket{e}$ is denoted as $\Gamma_e$, while the decay rate of state $\ket{r}$ is neglected.  The atomic coherences $\varrho_{gr}$ and $\varrho_{rg}$ decay with a rate $\gamma_{gr}$ which accounts for the sum of various dephasing processes, and is of the order of $\gamma_0 / 2 \pi=0.1$~MHz in absence of interactions \cite{han2015lensing}. The decay rates of other atomic coherences are assumed to be equal to $\Gamma_e/2$, i. e. they are predominantly determined by dissipation and additional dephasing rates are neglected in comparison.

%\bibliography{EITLensingEffect}
%
%
\end{document}